\documentclass[a4paper,11pt]{article}

\usepackage[left=2.5cm,right=2.5cm,top=2.5cm,bottom=2.5cm]{geometry}
\usepackage{graphicx,amssymb,amsmath,amsthm,mathrsfs,setspace,subcaption,cite,authblk,cancel,bbm,xcolor}

\usepackage[colorlinks=true,bookmarks=true]{hyperref}

\usepackage{mathtools}

\DeclarePairedDelimiterX\braket[2]{\langle}{\rangle}{#1 \delimsize\vert #2}

\hypersetup{citecolor=blue}
\newcommand{\dif}{\mathrm{d}}
\newcommand{\Eqref}[1]{(\ref{#1})}
\newcommand{\half}{\frac{1}{2}}

\newcommand{\brac}[1]{\left(#1 \right)}
\newcommand{\sbrac}[1]{\left[#1\right]}
\newcommand{\im}{\mathrm{i}}

\numberwithin{equation}{section}
 
\begin{document}

\title{Properties of the magnetic universe with positive cosmological constant}

\author{Yu-Xuan Toh\footnote{Email: yuxuantoh4@gmail.com}}

\author{Yun-Ten Chin\footnote{Email: yuntennn0109@gmail.com}}

\author{Ethan Quanrun Wu\footnote{Email: quanrun77@gmail.com}}

\author{Yen-Kheng Lim\footnote{Corresponding author, Email: yenkheng.lim@gmail.com, yenkheng.lim@xmu.edu.my} 
}

\affil{\normalsize{\textit{Department of Physics, Xiamen University Malaysia, 43900 Sepang, Malaysia}}}

\date{\normalsize{\today}}
\maketitle
 
\renewcommand\Authands{ and }

\begin{abstract}
  The properties of the Melvin-type spacetime with a positive cosmological constant $\Lambda$ in $d$-dimensional Einstein--Maxwell gravity is studied. The solution is parametrised in terms of the `de Sitter radius' $\ell\propto\Lambda^{-1/2}$ and the magnetic field parameter $\beta$, and they are warped products of the form $\mathbb{R}^{1,d-3}\times S^2$, where $\mathbb{R}^{1,d-3}$ is the $(d-2)$-dimensional Minkowski spacetime and $S^2$ is topologically a two-sphere which contains a conical singularity, whose nature depends on the product $\beta\ell$. In the limit $\ell\rightarrow\infty$, the $S^2$ decompactifies and the $d$-dimensional Melvin universe is recovered. The Freund--Rubin-type flux compactification model is shown to be another particular limit of this solution. We also calculate the flux and geodesics in this spacetime.
\end{abstract}

\section{Introduction} \label{sec_intro}

Under the mass-energy equivalence of General Relativity, a sufficiently strong magnetic field will carry enough energy to have its own gravity. An early notion of a configuration of magnetic field lines held together under its own gravity was envisioned as \emph{geons} by Wheeler \cite{Wheeler:1955zz}. Indeed, under Einstein--Maxwell gravity the electromagnetic field contributes to the stress-energy tensor in Einstein's equation, and an exact solution with zero cosmological constant describing a cylindrically-symmetric bundle of magnetic field lines was given by Melvin \cite{Melvin:1963qx} and is usually referred to as the \emph{Melvin universe}, though there are earlier works by Bonnor \cite{Bonnor:1954}, and Misra and Radhakrishna \cite{misra1962some}.

 {As a solution describing a bundle of axi-symmetric magnetic field lines, the Melvin solution in $d=4$ dimensions is a natural choice to locally describe the spacetime in the interior of a solenoid \cite{Davidson:1999fa,Zofka:2005av,Lim:2021kto}. With Einstein gravity being taken into account, the mass-energy of the magnetic field introduces curvature to the spacetime and therefore this model extends the elementary magnetostatic solenoid to include gravity. As such, they have been called \emph{cosmic solenoids} \cite{Davidson:1999fa} or \emph{relativistic solenoids} \cite{Zofka:2005av}.  In the typical non-relativistic solenoid the magnetic field is well known to be uniform and axi-symmetric. For the Melvin spacetime there are now effects of spacetime curvature and cosmological constant which distributes the magnetic field in a different way. When a negative cosmological constant $\Lambda$ is present, the additional attractive gravity it provides serves to confine the distribution of flux \cite{Kastor:2020wsm}. In this paper, where we are studying the positive $\Lambda$ counterpart, the repulsive force will serve to spread out the flux instead.
}

The properties of the $\Lambda=0$ solution was analysed by Melvin and other authors. In particular, geodesics and particle motion in the Melvin universe was studied in \cite{Melvin:1965zza,Melvin:1965zz}. The geometry and stability of the Melvin universe was studied by Thorne in \cite{Thorne:1965}. In \cite{Havrdova:2006gi}, Havrdov\'{a} and Krtou\v{s} showed that the Melvin solution can be obtained as a near-horizon limit of the charged C-metric. The generalisation of the Melvin universe (along with other black hole spacetimes in external magnetic fields) to higher dimensions was provided by Ortaggio \cite{Ortaggio:2004kr}.
The behaviour of scalar fields and perturbations on the Melvin spacetime was studied in \cite{Bini:2022xzk}. A further generalisation of the Melvin spacetime has been given in \cite{Golubtsova:2009esx,Bolokhov:2019jjf,Bolokhov:2019ktk}. Besides the Melvin universe, other cylindrically-symmetric spacetimes with radial magnetic fields have also been obtained and studied \cite{Vesely:2021jlc,Vesely:2022vws}. In Ref.~\cite{Cardoso:2024yrb} Cardoso and Nat\'{a}rio provided the scalar analogue of the Melvin universe, including the presence of the black hole.

The interplay between magnetic fields and gravity is also important in astrophysical contexts, as strong magnetic fields are present in accretion disks \cite{Bicak:2006hs,Jafari:2019brc}. Though typically there is a black hole at the centre of these disks, it is worthwhile to understand the behaviour of the magnetic field alone without the additional contribution of a black hole's gravity.\footnote{To add a black hole into a Melvin magnetic universe, one uses the Ernst solution \cite{Ernst:1976mzr}.} Besides black holes, strong magnetic fields may also play a role in the vicinity of neutron stars and magnetars \cite{Thompson:1995gw,Archibald:2013kla}, as well as influencing the cosmic expansion \cite{Barrow:2006ch,Durrer:2013pga}. Furthermore, as the Melvin solution represents a configuration of magnetic field in equilibrium under its own gravity. This may hint at the fact that magnetic fields may provide an outward force against the gravitational collapse of matter \cite{Tsagas:2020lal}. Seeing that a positive cosmological constant plays a leading, or simplest, contributor for dark energy \cite{Padmanabhan:2002ji,Peebles:2002gy,Li:2012dt}, it is worthwhile to extend the study of Melvin-type magnetic universes with positive cosmological constant.

The extension of the Melvin spacetime to non-zero cosmological constant $\Lambda$ was obtained by Astorino via a generalisation of the Ernst potentials \cite{Astorino:2012zm}. Another approach to include a cosmological constant to the Melvin spacetime was given by \v{Z}ofka in \cite{Zofka:2019yfa}. The solution can also be derived as a near-horizon limit of the (Anti-)de Sitter C-metric \cite{Lim:2018vbq}. The properties of the case $\Lambda<0$ was studied in \cite{Kastor:2020wsm,Lim:2021kto}. A notable feature of the $\Lambda<0$ case is that it is asymptotic to Anti-de Sitter space in Poincar\'{e} slicing, whereas the $\Lambda=0$ and $\Lambda>0$ solutions are neither asymptotically-flat nor de Sitter, respectively. More recently, there are further related works found in \cite{Bouzenada:2024ryh,Castro:2024ayd,Ahmed:2025ohc}.

 {
For gravity in dimensions $d\geq4$, we will see that it describes a gravitational model with two compactified extra dimensions. In recent decades there has been interest in models of gravity such as string theory or braneworld gravity \cite{Freund:1980xh,Randall:1999ee,Randall:1999vf}, it is important to have a mechanism that compactifies or `hides' the extra dimension in order to recover the four-dimensional macroscopic universe that we observe. A widely-used model for this is the Freund--Rubin \cite{Freund:1980xh} flux compactification model, in which a $q$-form flux serves to compactify a theory with $q$ extra dimensions. We will show that the Melvin universe with a positive cosmological constant is closely related to flux compactification, specifically in the case $q=2$ wherein our $q=2$-form is the Maxwell field.}

 {In light of the various physical systems that the Melvin universe may describe, along with the fact that it's one of the few known exact solutions involving a non-trivial magnetic and gravitational field, it is worth seeking a complete understanding on this solution. Particularly how magnetic fields induce spacetime curvature and conversely how a curved spacetime influences a configuration of magnetic fields.} To our knowledge, the properties of the $\Lambda>0$ solution has yet to be studied closely, and this will be the main task of this paper. The spacetime is characterised by two parameters, namely the magnetic field strength $\beta$ and the curvature scale of the cosmological constant $\ell\propto\Lambda^{-1/2}$. The spacetime has the topology of the product between a Minkowski spacetime and a distorted sphere carrying a conical singularity. As noted before in \cite{Astorino:2012zm,Lim:2018vbq}, the $\Lambda\neq0$ Melvin solution can be obtained as a double-Wick rotation of the Reissner--Nordstr\"{o}m black hole with planar horizon, and a coordinate transformation can be applied to cast it into Melvin's familiar form. Here, we find it convenient to remain in the Reissner--Nordstr\"{o}m-like coordinates. In the $\Lambda>0$ case, the solution also contains the Freund--Rubin-type compactification model \cite{Freund:1980xh,Bousso:2002fi} as a particular limit. Recall that these Freund--Rubin-type models describe spacetimes with $q$ compact dimensions stabilised by a $q$-form flux. The flux in the Melvin solution is just the Faraday tensor, hence $q=2$ in our case, and we will show how a limiting procedure recovers the Freund--Rubin model with two compact dimensions.

The rest of this paper is organised as follows. In Sec.~\ref{sec_eom} we review the equations of motion and the Melvin universe solution with positive cosmological constant. The physical and geometrical properties are studied in Sec.~\ref{sec_GeomProp}.  In Sec.~\ref{sec_flux}, we analyse the magnetic flux of the solution, followed by a study of geodesics in Sec.~\ref{sec_geodesics}. Conclusions and closing remarks are given in Sec.~\ref{sec_conclusion}. Our convention for Lorentzian signature is $(-,+,\ldots,+)$ and we work in geometric units where $c=G=1$.

\section{The Melvin solution with non-zero cosmological constant} \label{sec_eom}

Our framework to build the Melvin solution is a $d$-dimensional Einstein--Maxwell theory with a cosmological constant. The spacetime $(M,g_{\mu\nu})$ is of dimension $d$, and the electromagnetic field is described by the Faraday tensor $F=\half F_{\mu\nu}\dif x^\mu\wedge\dif x^\nu=\dif A$, which is the exterior derivative of the 1-form potential $A=A_\mu\dif x^\mu$. The components of the Faraday tensor are computed using $F_{\mu\nu}=\nabla_\mu A_\nu-\nabla_\nu A_\mu$.

\subsection{Action and equations of motion}

The action describing this theory is
\begin{align}
 I=\frac{1}{16\pi}\int_M\dif^dx\sqrt{|\det g|}\brac{R-F^2-2\Lambda},
\end{align}
where $R$ is the Ricci scalar, $F^2=F_{\mu\nu}F^{\mu\nu}$ is the self-contraction of the Faraday tensor, and $\Lambda$ is the cosmological constant. Extremising the action with respect to the metric and potential leads to the Einstein--Maxwell equations
\begin{subequations} \label{EME}
\begin{align}
 R_{\mu\nu}&=\frac{2\Lambda}{d-2}g_{\mu\nu}+2F_{\mu\lambda}{F_\nu}^\lambda-\frac{1}{d-2}F^2g_{\mu\nu},\\
 \nabla_\lambda F^{\lambda\mu}&=0.
\end{align}
\end{subequations}
We now show that the Melvin-type solution with non-zero cosmological constant can be obtained by applying two Wick rotations to the (Anti-)de Sitter--Reissner--Nordstr\"{o}m solution,
\begin{subequations}
\begin{align}
 \dif s^2&=-f(r)\dif\tau^2+\frac{\dif r^2}{f(r)}+\frac{r^2}{\lambda^2}\tilde{\gamma}_{ab}\dif x^a\dif x^b,\\
 A&=\brac{\Phi+\sqrt{\frac{d-2}{2(d-3)}}\;\frac{q}{r^{d-3}}}\dif\tau,\\
 f(r)&=\frac{k}{\lambda^2}-\frac{\mu}{r^{d-3}}-\frac{r^2}{\ell^2}+\frac{q^2}{r^{2(d-3)}},\quad \ell^2=\frac{(d-1)(d-2)}{2\Lambda}.
\end{align}
\end{subequations}
where $\mu$, and $q$ are constants. In the solution for the potential, $\Phi$ is an arbitrary gauge constant which we later fix to convenient values. The $(d-2)$-dimensional metric $\tilde{\gamma}_{ab}$ is a maximally-symmetric metric with curvature scale $\lambda$. The discrete parameter $k=\pm1, 0$ specifies whether the curvature is positive, negative, or zero. This solution describes a charged black hole with spherical, planar, or hyperbolic horizon for $k=1$, $k=0$, or $k=-1$, respectively. To obtain the Melvin solution, we consider the planar case $k=0$ where $\tilde{\gamma}_{ab}$ is simply the flat Euclidean space $\mathbb{R}^{d-2}$. Then the metric is
\begin{align*}
 \dif s^2&=-f(r)\dif\tau^2+\frac{\dif r^2}{f(r)}+\frac{r^2}{\lambda^2}\brac{\dif y^2+\dif x_1^2+\ldots+\dif x_{d-3}^2}.
\end{align*}
Performing the transformations
\begin{align}
 \tau\rightarrow\im\psi,\quad y\rightarrow\im t,\quad \mu\rightarrow-\nu,\quad q\rightarrow\im\beta,\quad\Phi\rightarrow-\im\chi_0,
\end{align}
where $t$ and $\psi$ are the new time and spatial coordinates after Wick rotating $y$ and $\tau$, along with $\nu$, $q$, and $\chi_0$ being new constant parameters. The solution then becomes
\begin{subequations} \label{dSMel_soln}
\begin{align}
 \dif s^2&=\frac{r^2}{\lambda^2}\eta_{ab}\dif x^a\dif x^b+\frac{\dif r^2}{f(r)}+f(r)\dif\psi^2,\\
 A&=\brac{-\sqrt{\frac{d-2}{2(d-3)}}\frac{\beta}{r^{d-3}}+\chi_0}\dif\psi,\label{GaugePotential}\\
 f(r)&=\frac{\nu}{r^{d-3}}-\frac{r^2}{\ell^2}-\frac{\beta^2}{r^{2(d-3)}},\quad \ell^2=\frac{(d-1)(d-2)}{2\Lambda}, \label{f_eqn}
\end{align}
\end{subequations}
where $\eta_{ab}\dif x^a\dif x^b=-\dif t^2+\dif x_1^2+\ldots+\dif x_{d-3}^2$ denotes the $(d-2)$-dimensional Minkowski spacetime $\mathbb{R}^{1,d-3}$ with its coordinates indexed by lowercase Latin indices $x^a$. The constant $\lambda$ remains an arbitrary constant setting the length scale of the coordinates. After the transformation, the field $F_{\mu\nu}=\nabla_\mu A_\nu-\nabla_\nu A_\mu$ is now purely magnetic. One can directly verify that the solution \Eqref{dSMel_soln} still solves the Einstein--Maxwell equations \Eqref{EME}. In the case of vanishing field $\beta=0$, and negative cosmological constant $\Lambda<0$, the spacetime is the Horowitz--Myers soliton \cite{Horowitz:1998ha}.

\subsection{Transformation to `cylindrical' coordinates}

We now show that \Eqref{dSMel_soln} is indeed the  Melvin magnetic universe, generalised to include a cosmological constant. As mentioned in Sec.~\ref{sec_intro}, such a solution was already found in $d=4$ by Astorino \cite{Astorino:2012zm}. Therefore we have to show that the solution \Eqref{dSMel_soln} is equivalent to the one in \cite{Astorino:2012zm} upon coordinate transformation.

To cast the solution in Astorino's form, we introduce the transformation
\begin{align}
 \frac{r}{\lambda}&=\brac{1+\frac{1}{4}B^2\rho^2}^{\frac{1}{d-3}},\quad \psi=\frac{2(d-3)}{\lambda B^2}\phi, \label{AstorinoCoord}
\end{align}
where $\rho\geq0$ is the `cylindrical' radial coordinate and $B$ is a constant with dimension of inverse length. In these new coordinates the solution becomes
\begin{subequations}
\begin{align}
 \dif s^2&=H^{\frac{2}{d-3}}\brac{\eta_{ab}\dif x^a\dif x^b+\frac{\dif\rho^2}{Q}}+\frac{Q\rho^2}{H^2}\dif\phi^2,\\
 A&=\brac{-\sqrt{\frac{d-2}{2(d-3)}}\,\frac{2}{BH}+\frac{2(d-3)}{\lambda B^2}\chi_0}\dif\phi,\\
 Q&=\frac{4}{B^4\rho^2}\brac{\frac{(d-3)^2\nu}{\lambda^{d-1}}H-\frac{(d-3)^2\beta^2}{\lambda^{2(d-2)}}-\frac{(d-3)^2}{\ell^2}H^{\frac{2(d-2)}{(d-3)}}},\quad H=1+\frac{1}{4}B^2\rho^2.
\end{align}
\end{subequations}
Choosing the arbitrary constant to be $\chi_0=\sqrt{\frac{d-2}{2(d-3)}}\,\frac{\lambda B}{d-3}$, the gauge potential then becomes
\begin{align}
 A=\sqrt{\frac{d-2}{2(d-3)}}\,\frac{B\rho^2}{H}\dif\phi,
\end{align}
which is more explicitly similar to the potential as appearing in \cite{Astorino:2012zm}.\footnote{Specifically, it is given in the paragraph following Eq.~(4.7) in \cite{Astorino:2012zm}.}

However, for general $\nu$, we have $\lim_{\rho\rightarrow0}Q=\infty$, so the metric is singular. To remove this singularity we first let
\begin{align}
 \beta=\frac{\lambda^{d-2}B}{d-3}, \label{Btransform}
\end{align}
and then fix $\nu$ to satisfy
\begin{align}
 \frac{\nu}{\lambda^{d-3}}=\frac{\lambda^2B^2}{(d-3)^2}+\frac{\lambda^2}{\ell^2}. \label{fix_nu}
\end{align}
The metric and gauge potential now take the form
\begin{subequations} \label{Astorino_soln}
\begin{align}
 \dif s^2&=H^{\frac{2}{d-3}}\brac{\eta_{ab}\dif x^a\dif x^b+\frac{\dif\rho^2}{Q}}+\frac{Q\rho^2}{H^2}\dif\phi^2,\\
 A&=\sqrt{\frac{d-2}{2(d-3)}}\,\frac{B\rho^2}{H}\dif\phi,\\
 Q&=\frac{4}{B^4\rho^2}\sbrac{\brac{B^2+\frac{(d-3)^2}{\ell^2}}H-B^2-\frac{(d-3)^2}{\ell^2}H^{\frac{2(d-2)}{d-3}}},\quad H=1+\frac{1}{4}B^2\rho^2.
\end{align}
\end{subequations}
We see that that in the limit of zero cosmological constant, $\ell\rightarrow\infty$, we have $Q\rightarrow1$ and we recover the $d$-dimensional Melvin universe \cite{Ortaggio:2004kr}. In the case $d=4$, along with
\begin{align}
 \phi=\frac{B\,\bar{\varphi}}{\sqrt{B^4-B^2\Lambda}},\quad \left.\ell^2\right|_{d=4}=\frac{3}{\Lambda},
\end{align}
we precisely recover Eq.~(4.7) of \cite{Astorino:2012zm}. This shows that the solution \Eqref{dSMel_soln} is equivalent to the Melvin-like solution first obtained in the form \Eqref{Astorino_soln} under a coordinate transformation. In the following sections we will continue using the form \Eqref{dSMel_soln}, as its Schwarzschild/Reissner--Nordstr\"{o}m-like appearence is more convenient for our purposes.

\subsection{Limit to a flux compactification model} \label{subsec_FluxComp}

In this subsection we wish to show that the solution \Eqref{dSMel_soln} also contains the Freund--Rubin-type flux compactification model \cite{Freund:1980xh,Bousso:2002fi} as a particular limit. To see this we write $f(r)$ as given in Eq.~\Eqref{f_eqn} as
\begin{align}
 f(r)=\frac{1}{r^{2(d-3)}}\brac{\nu r^{d-3}-\frac{1}{\ell^2}r^{2(d-2)}-\beta^2}=\frac{1}{r^{2(d-3)}}P(r).
\end{align}
Note that the expression in the parenthesis is a polynomial $P(r)$ of degree $2(d-2)$ consisting of three monomials. By applying the Descartes Rule of Signs, this polynomial has at most two positive roots, which we denote as $\lambda$ and $r_0$. Let $r_j$, $j=1,\ldots,2(d-3)$ be the remaining negative or complex conjugate roots and write $P(r)$ in terms of its root factorisation,
\begin{align}
 P(r)=\frac{1}{\ell^2}(r-\lambda)(r_0-r)\prod_{j=1}^{2(d-3)}(r-r_j).
\end{align}

To obtain the Freund--Rubin model, we take the limit $r_0\rightarrow\lambda$ while simultaneously scaling up the $r$-coordinate so that the domain $\lambda\leq r\leq r_0$ remains finite. Therefore we let
\begin{align}
 r_0=\lambda+2\epsilon b,\quad r=\lambda\epsilon(x+b),
\end{align}
where $x$ is the new coordinate centred between $r_0$ and $\lambda$, $b$ is an arbitrary length scale, and $\epsilon$ is a dimensionless parameter which will be later taken to zero. In terms of these new parameters and coordinate, the polynomial $P$ takes the form
\begin{align*}
 P(x)=\epsilon^2\brac{b^2-x^2}Q(x),\quad Q(x)=\frac{1}{\ell^2}\prod_{j=1}^{2(d-3)}(\lambda-r_j+\epsilon(x+b)),
\end{align*}
where $Q$ tends to a constant $C$ in the limit $\epsilon\rightarrow0$. The metric and gauge potential becomes
\begin{subequations}
\begin{align}
 \dif s^2&=\frac{\epsilon^2\brac{b^2-x^2}Q}{\sbrac{\lambda+\epsilon(x+b)}^{2(d-3)}}\dif\psi^2+\frac{\sbrac{\lambda+\epsilon(x+b)}^{2(d-3)}\dif x^2}{\brac{b^2-x^2}Q}+\frac{\brac{\lambda+\epsilon(x+b)}^2}{\lambda^2}\eta_{ab}\dif x^a\dif x^b,\\
 A&=\brac{-\sqrt{\frac{d-2}{2(d-3)}}\frac{\epsilon}{\sbrac{\lambda+\epsilon(x+b)}^{d-3}}+\chi_0}\dif\psi.
\end{align}
\end{subequations}
Here $Q$ tends to a constant $C$ in the limit $\epsilon\rightarrow0$. If we choose
\begin{align*}
 \chi_0=-\sqrt{\frac{d-2}{2(d-3)}}\,\frac{1}{\lambda^{d-3}},
\end{align*}
and $\lambda$ to satisfy $b^2C/\lambda^{2(d-3)}=1$, along with rescaling the angular coordinate to $\psi=\frac{b}{\epsilon}\varphi$, the limit $\epsilon\rightarrow0$ gives
\begin{subequations}
\begin{align}
 \dif s^2&=\brac{1-\frac{x^2}{b^2}}\dif\varphi^2+\frac{\dif x^2}{1-x^2/b^2}+\eta_{ab}\dif x^a\dif x^b,\\
 A&=qx\dif\varphi,
\end{align}
\end{subequations}
where $q=\sqrt{\half(d-2)(d-3)}\;\frac{\beta}{\lambda^{d-3}}$. We see that this is a Freund--Rubin type flux compactification $\mathbb{R}^{1,d-3}\times S^2$ \cite{Freund:1980xh,Bousso:2002fi} specifically with a $(d-2)$-dimensional Minkowsksi spacetime $\mathbb{R}^{1,d-3}$ compactified over a two-sphere $S^2$. The flux is the two-form $F=\dif A=q\,\dif x\wedge\dif\varphi$.

\section{Geometrical properties} \label{sec_GeomProp}

 {The solution \Eqref{dSMel_soln} describes a warped product between $(d-2)$-dimensional Minkowski spacetime with two `extra' dimensions. In this section, we will show below that in order to avoid a naked curvature singularity it is neccessary that a magnetic field is present and the `extra' dimensions are compact. The geometry of this compact space is that of a distorted sphere, and we will investigate its geometrical properties. In a sense, the solution describes a form of Kaluza--Klein-type spacetime with two compact dimensions, where we have already seen in the previous section that it contains the Freund--Rubin model as a limit. Unlike the Freund--Rubin limit, the geometry of the the compactified dimension is not spherically symmetric, and contains a conical singularity. The  conical singularity itself may be important if this spacetime were to be interpreted as a braneworld scenario, as the conical singularity is interpreted as the matter source for the branes and play a role in their gravitational thermodynamics \cite{Mukohyama:2005yw,Kinoshita:2006eld,Kinoshita:2007ci}.
}


\subsection{Parameter and coordinate domains} \label{subsec_ParamSpace}

Let us inspect again the function
\begin{align}
 f=\frac{\nu}{r^{d-3}}-\frac{r^2}{\ell^2}-\frac{\beta^2}{r^{2(d-3)}}.
\end{align}
We are mainly interested in the case of positive cosmological constant, $\ell^2>0$. Clearly, we require $f>0$ for the metric to carry a Lorentzian signature. This then requires $\nu>0$. If the magnetic field is absent ($\beta=0$), then the Lorentzian region lies in the domain
\begin{align*}
 0<r<\brac{\nu\ell^2}^{1/(d-1)}.
\end{align*}
This domain contains a naked curvature singularity at $r\rightarrow0$. This can be seen, say, from the Kretschmann scalar
\begin{align}
 R_{\mu\nu\rho\sigma}R^{\mu\nu\rho\sigma}&=f''^2+\frac{2(d-2)}{r^2}f'^2-\frac{2(d-2)(d-3)}{r^4}f^2,
\end{align}
which diverges as $r\rightarrow0$. To avoid this curvature singularity, we require $\beta$ to be non-zero. Then the presence of the term $-\beta^2/r^{2(d-3)}$ in $f$ will have the spacetime terminate at $f=0$ with a finite root $r>0$ before a singularity is encountered. At $f=0$, we have $g_{\psi\psi}=0$ which indicates the `tip' of the spacetime, and points with $r=0$ are no longer part of the manifold.

 {These considerations constrain $f$ to be positive and have two roots and the direction parametrised by coordinate $r$ is compact. The finiteness of this domain has two non-trivial physical consequences. As a model of higher-dimensional gravity, this means that the solution is a Kaluza--Klein-type spacetime with two compact dimensions. On the other hand, if viewed as a local description of a solenoid interior, the finiteness of $r$ means that one cannot have an interior with an arbitrarily large cross section, as the space eventually curves in and closes on itself. This is different from a solenoid with zero or positive cosmological constant, as $r$ is unbounded and so their cross sections can be arbitrarily large.}

\begin{figure}
 \centering
 \includegraphics{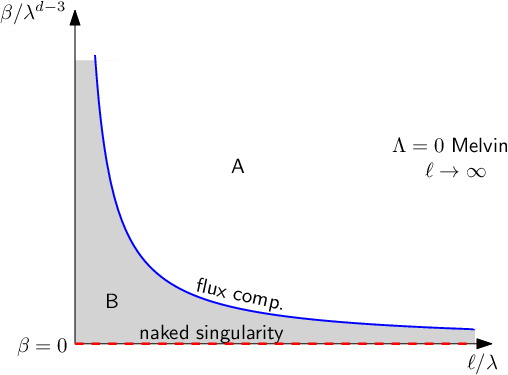}
 \caption{The $(\ell,\beta)$-parameter space for the Melvin universe with positive cosmological constant. \textsf{Case A} is the unshaded domain $\beta\ell>\sqrt{\frac{d-1}{d-3}}\lambda^{d-4}$, where $\lambda<r_0$, and \textsf{Case B} is the shaded region $\beta\ell>\sqrt{\frac{d-1}{d-3}}\lambda^{d-4}$ where $\lambda>r_0$. The dashed red line $\beta=0$ corresponds to $r_0=0<r<\lambda$ which is a naked singularity. The blue curve $\beta\ell=\sqrt{\frac{d-1}{d-3}}\lambda^{d-4}$ is the limit $r_0\rightarrow\lambda$, which was shown in Sec.~\ref{subsec_FluxComp} to be the limit to the Freund--Rubin flux compactification model.}
 \label{fig_ParamSpace}
\end{figure}

We can fix our coordinate system in terms of the `tip' where $f=0$. In other words, we fix $\nu$ following Eq.~\Eqref{fix_nu}, so that the function $f$ becomes
\begin{align}\label{function fr}
 f(r)=\brac{\frac{\beta^2}{\lambda^{2(d-3)}}+\frac{\lambda^2}{\ell^2}}\frac{\lambda^{d-3}}{r^{d-3}}-\frac{r^2}{\ell^2}-\frac{\beta^2}{r^{2(d-3)}}.
\end{align}
This means $\lambda$ is a root of $f$, and the solution caps off there. This essentially fixes $\lambda$ as the arbitrary length scale of the system,  {and the Melvin solution is described by two free physical parameters, namely
\begin{align}
 \ell=\sqrt{\frac{(d-1)(d-2)}{2\Lambda}},\quad\mbox{and}\quad \beta,
\end{align}
which, to reiterate, are the cosmological constant and magnetic field strength, respectively. They can be expressed in units of $\lambda$. We will see in the rest of this paper how $\ell$ and $\beta$ affect the geometry and flux configuration of the spacetime.
}
Checking the derivative at $r=\lambda$, we find
\begin{align}\label{f prime lambda}
 f'(\lambda)=\frac{(d-3)\beta^2}{\lambda^{2d-5}}-\frac{(d-1)\lambda}{\ell^2}\left\{\begin{array}{cc}
                           >0 & \mbox{ if }\frac{1}{\ell^2}<\frac{(d-3)\beta^2}{(d-1)\lambda^{2(d-2)}},\\
                           <0 & \mbox{ if }\frac{1}{\ell^2}>\frac{(d-3)\beta^2}{(d-1)\lambda^{2(d-2)}},\end{array}\right.
\end{align}
if we assume there exists another real root $r_0$ besides $\lambda$. Then the domains of positive $f$ consist of two distinct classes
\begin{subequations}
\begin{align}
 \mbox{Case A:}\quad\lambda<r<r_0&\quad\mbox{ if }\quad \beta\ell>\eta,\label{lambda_left}\\
 \mbox{Case B:}\quad r_0<r<\lambda&\quad\mbox{ if }\quad \beta\ell<\eta,\label{lambda_right}
\end{align}
\end{subequations}
where $\eta$ is a critical value given by
\begin{align}
 \eta=\sqrt{\frac{d-1}{d-3}}\,\lambda^{d-2}.\label{def_eta}
\end{align}
In Case A, the root $r_0$ tends to infinity as $\ell^2\rightarrow\infty$, whereas in Case B, $r_0$ approaches zero as $\beta\rightarrow0$. The critical case $\beta\ell=\eta$ corresponds to $\lambda$ coinciding with $r_0$, which is the Freund--Rubin limit discussed in Sec.~\ref{subsec_FluxComp}. The parameter space plotted as a $(\ell,\beta)$-plane is sketched in Fig.~\ref{fig_ParamSpace}. On this plane, our two cases A and B are the unshaded and shaded region, respectively. They are separated by a common boundary in the shape of a hyperbola $\beta\ell=\eta$ corresponding to the Freund--Rubin solution.

\begin{figure}
 \centering
 \begin{subfigure}[b]{0.49\textwidth}
    \centering
    \includegraphics[width=\textwidth]{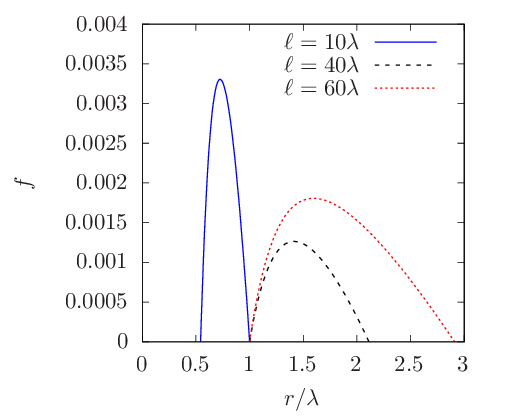}
    \caption{}
    \label{fig_froots_fixedbeta_dS}
  \end{subfigure}
  \begin{subfigure}[b]{0.49\textwidth}
    \centering
    \includegraphics[width=\textwidth]{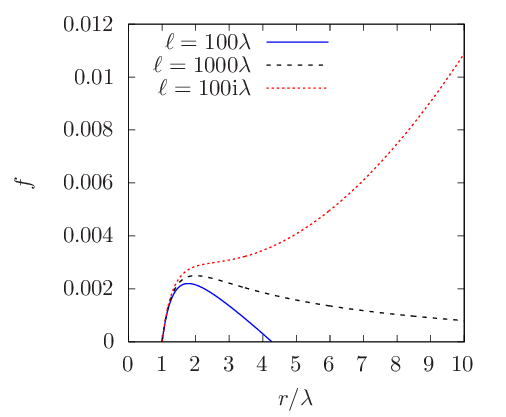}
    \caption{}
    \label{fig_froots_fixedbeta_all}
  \end{subfigure}
  \caption{Plots of $f(r)$ vs $r$ for $d=4$, $\beta=0.1\lambda$ at various $\ell$. In Fig.~\ref{fig_froots_fixedbeta_all}, the imaginary value $\ell=100\im\lambda$ means a negative cosmological constant, $\Lambda=\frac{3}{\ell^2}<0$.}
  \label{fig_froots_fixedbeta}
\end{figure}

For some numerical examples, let us consider $d=4$ with fixed $\beta=0.1\lambda$. The graphs of $f(r)$ vs $r$ for $\ell=10\lambda$ (Case B), $40\lambda$ (Case A), and $60\lambda$ (Case A) are plotted in Fig.~\ref{fig_froots_fixedbeta_dS}. The second roots are $r_0=0.5437\lambda$, $2.1124\lambda$, and $2.9094\lambda$, respectively.  {Confirming the expectations of the above calculations, positive $f$ occurs in a finite domain of $r$, demonstrating that the $(r,\psi)$-section is compact thus showing the non-existence of a $\Lambda>0$ solenoid with arbitrarily large cross section. In Fig.~\ref{fig_froots_fixedbeta_all} we plot the graphs for $\ell=100\lambda$, $1000\lambda$, and $100\im\lambda$. As expected the space becomes decompactified as the cosmological constant approaches zero ($\ell\propto\Lambda^{-1/2}\rightarrow\infty$), along with the $\Lambda<0$ case still being non-compact.}

In the latter we formally write imaginary $\ell$ to mean a negative cosmological constant, $\ell^2<0$. Note that as $|\ell|$ gets larger, we approach the limit of zero cosmological constant, and $r_0\rightarrow\infty$ giving the limit of the $\Lambda=0$ Melvin universe. For imaginary $\ell$ (or $\Lambda<0$), the spacetime is asymptotically Anti-de Sitter \cite{Kastor:2020wsm}.

\begin{figure}
 \begin{subfigure}[b]{0.49\textwidth}
    \centering
    \includegraphics[width=\textwidth]{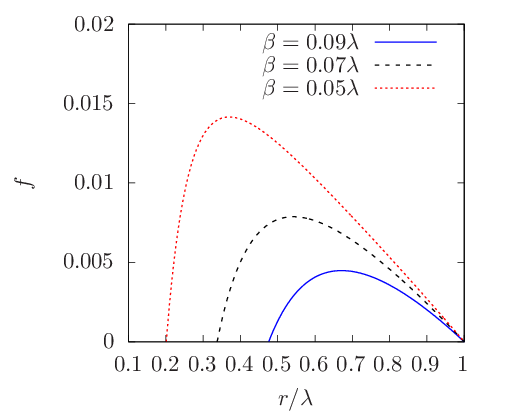}
    \caption{$\ell=10\lambda$.}
    \label{fig_froots_fixedl_lambdaleft}
  \end{subfigure}
  \begin{subfigure}[b]{0.49\textwidth}
    \centering
    \includegraphics[width=\textwidth]{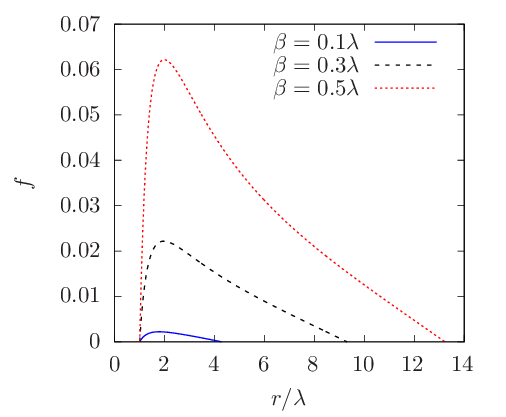}
    \caption{$\ell=100\lambda$.}
    \label{fig_froots_fixedl_lambdaright}
  \end{subfigure}
  \begin{subfigure}[b]{0.49\textwidth}
    \centering
    \includegraphics[width=\textwidth]{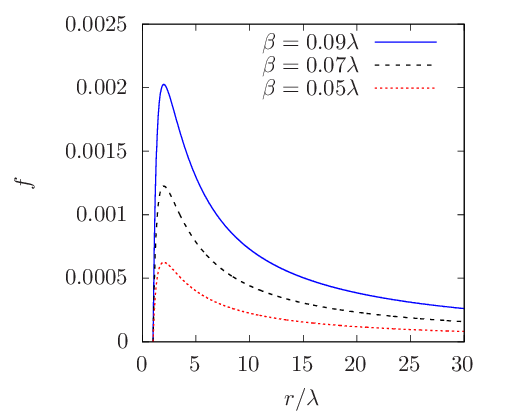}
    \caption{$\ell\rightarrow\infty$.}
    \label{fig_froots_fixedl_Mink}
  \end{subfigure}
  \begin{subfigure}[b]{0.49\textwidth}
    \centering
    \includegraphics[width=\textwidth]{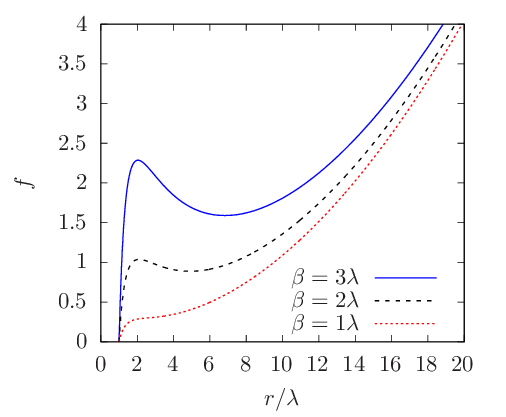}
    \caption{$\ell=10\im\lambda$.}
    \label{fig_froots_fixedl_AdS}
  \end{subfigure}
  \caption{Plots of $f(r)$ vs $r$ for $d=4$, (a) $\ell=10\lambda$, (b) $\ell=100\lambda$, (c) $\ell\rightarrow\infty$, and $\ell=10\im\lambda$}
  \label{fig_froots_fixedl}
\end{figure}

In Fig.~\ref{fig_froots_fixedl} we plot the function for fixed $\ell$ to see how the parameter $\beta$ affects the shape of $f(r)$. In Fig.~\ref{fig_froots_fixedl_lambdaleft} we consider $d=4$, $\ell=10\lambda$ and in Fig.~\ref{fig_froots_fixedl_lambdaright} we plotted for $\ell=100\lambda$. We see that as $\beta$ is adjusted below (resp. above) from the critical value $\beta\ell=\eta$, the second roots $r_0$ below (resp. above) $\lambda$.  {Note that these pictures are coordinate-dependent, but the compact topology is represented by the finiteness of the domain for $f>0$. A more meaningful physical picture may be obtained by calculating proper lengths and embedding diagrams in Sec.~\ref{subsec_conical}.
}

For comparison we have also included the zero cosmological constant (Fig.~\ref{fig_froots_fixedl_Mink}) and negative cosmological constant (Fig.~\ref{fig_froots_fixedl_AdS}) cases.  {In the former, we have the well known behaviour that $f$ asymptotically vanishes in the $\Lambda=0$ case, and the maximum of $f$ increase with the field strength $\beta$. The $\Lambda=0$ case is not asymptotically flat. Similarly for $\Lambda<0$, there is also a local maximum of $f$ which increases with field strength, followed by a local minimum, after which $f$ grows and scales like $r^2$, showing that the spacetime is asymptotically Anti-de Sitter in Poincar\'{e} slicing.}

\subsection{Geometry of the \texorpdfstring{$(r,\psi)$}{(r,psi)} section} \label{subsec_conical}

The geometry of the $(r,\psi)$-part of the spacetime can be inferred by considering circles of constant $x^a$ and $r$. We place the centre of the circle at $r=\lambda$, its radius at coordinate $r$ is
\begin{align}
 \mathcal{R}&=\pm\int_\lambda^r\sqrt{g_{rr}}\dif r'=\pm\int_\lambda^r\frac{\dif r'}{\sqrt{f(r')}}, \quad \beta^2\ell^2\gtrless\eta, \label{define_Rcal}
\end{align}
where the upper and lower signs correspond to cases \Eqref{lambda_left} and \Eqref{lambda_right}, respectively. The proper circumference of the circles are given by
\begin{align}
 \mathcal{C}&=\int_0^{\Delta\psi}\sqrt{g_{\psi\psi}}\dif\psi=\Delta\psi\sqrt{f(r)}.
\end{align}

Next we check for the presence of conical singularities at the poles $r=\lambda$ and $r=r_0$. To do this we draw small circles around the poles and check the ratios
\begin{align*}
 \gamma_\lambda=\lim_{r\rightarrow\lambda}\frac{\mathcal{C}}{\mathcal{R}},\quad \gamma_0=\lim_{r\rightarrow r_0}\frac{\mathcal{C}}{\mathcal{R}}
\end{align*}
in the neighbourhood close to each pole. There will be a conical deficit if $\gamma<2\pi$, conical excess if $\gamma>2\pi$, and no conical singularity if $\gamma=2\pi$. Computing for $r=\lambda$ and $r=r_0$, we have
\begin{align}
 \gamma_\lambda=\half\Delta\psi\left|f'(\lambda)\right|,\quad \gamma_0=\half\Delta\psi\left|f'(r_0)\right|,
\end{align}
where $\Delta\psi$ is the periodicity of the coordinate $\psi$. Let us remove the conical singularity at $r=\lambda$ by choosing the periodicty
\begin{align}
 \Delta\psi=\frac{4\pi}{|f'(\lambda)|}=\frac{4\pi}{\left|\frac{(d-3)\beta^2}{\lambda^{2d-5}}-\frac{(d-1)\lambda}{\ell^2} \right|}. \label{set_delta_psi}
\end{align}
This leaves a conical singularity at $r_0$ with
\begin{align}
 \gamma_0=2\pi\frac{|f'(r_0)|}{|f'(\lambda)|}.
\end{align}
In general, it is not possible to fix the parameters such that $|f'(r_0)|=|f'(\lambda)|$, therefore at least one conical singularity has to be present. In particular {it can be shown that} Case A solutions ($\beta\ell>\eta$) have conical deficit while Case B solutions ($\beta\ell<\eta$) have conical excess.


In the $d=4$ case, this was shown to be inherited from the conical singularity of the de Sitter C-metric \cite{Lim:2018vbq}. Only in the flux compactification limit ($\beta\ell\rightarrow\eta$) do we have $\gamma_0=2\pi$. Indeed, the Freund--Rubin-type  solutions consists of round (hence, regular) spheres $S^2$ as the compactified space.

\begin{figure}
 \centering
 \includegraphics{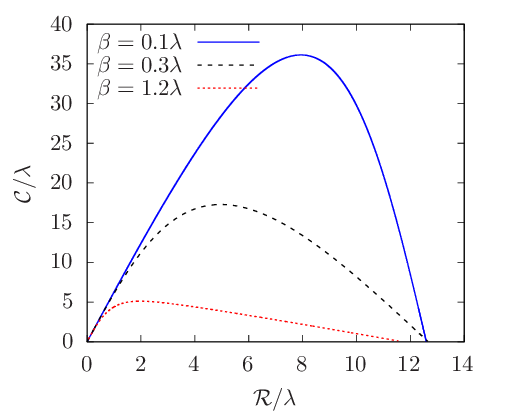}
 \caption{Plots of circumference $\mathcal{C}$ vs radius $\mathcal{R}$ of constant $(r,x^a)$ circles about $\lambda$ for $d=4$, $\ell=10\lambda$. As we have chosen the periodicity of $\psi$ to remove the conical singularity at $\lambda$, the tangents of each curve near $\mathcal{R}\sim 0$ have slopes $2\pi$.}
 \label{fig_RC}
\end{figure}

With this periodicity of $\psi$ chosen, we can now establish $\mathcal{C}$ vs $\mathcal{R}$ of the constant-$(r,x^a)$ circles centred at $r=\lambda$. This is shown in Fig.~\ref{fig_RC} for $\ell=10\lambda$ and various $\beta$. Because we removed the conical singularity at $r=\lambda$, small circles around $r=\lambda$ will have $\mathcal{C}/\mathcal{R}\simeq2\pi$. Therefore the tangents of the graphs near $\mathcal{R}=0$ have slopes $2\pi$ in Fig.~\ref{fig_RC}. Starting from $\mathcal{R}=0$ ($r=\lambda$), we see that the circles have increasing circumferences until some maximum value, before decreasing again towards zero as $r=r_0$ is approached. This is in contrast with the $\Lambda=0$ Melvin, where circles only asymptotically approach zero circumference infinitely far away. In Fig.~\ref{fig_RC}, the solid blue curve corresponds to $\beta\ell<\eta$ (Case B) and hence the second intersection with the horizontal axis occurs at a slope whose magnitude is greater than $2\pi$ reflecting the conical excess. The other two curves (black dashed and red dotted) have $\beta\ell>\eta$ (Case A), and their second intersections occurs at  slopes with magnitudes less than $2\pi$ demonstrating their conical deficit.

\begin{figure}
 \begin{subfigure}[b]{0.49\textwidth}
    \centering
    \includegraphics[width=\textwidth]{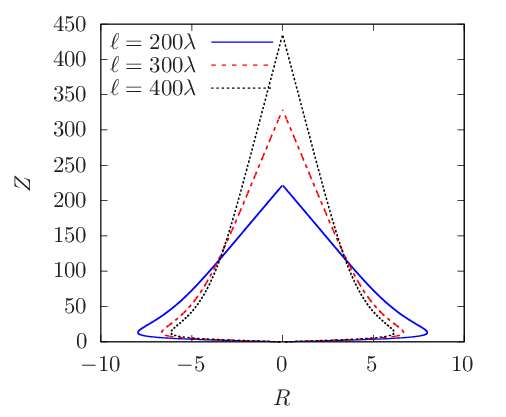}
    \caption{$\beta=0.2\lambda$.}
    \label{fig_embedding_Varyl}
  \end{subfigure}
  \begin{subfigure}[b]{0.49\textwidth}
    \centering
    \includegraphics[width=\textwidth]{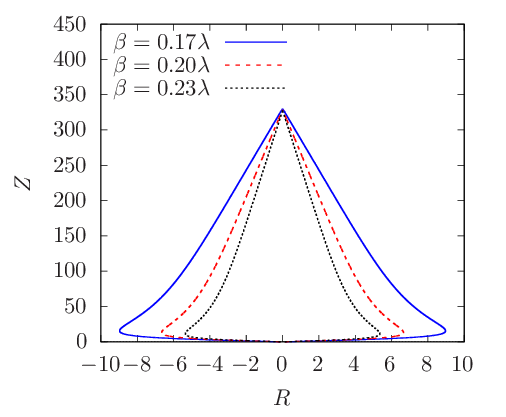}
    \caption{$\ell=300\lambda$.}
    \label{fig_embedding_Varybeta}
  \end{subfigure}
 \caption{Embedding diagram for $d=4$ in various $\ell$ and $\beta$. In Fig.~\ref{fig_embedding_Varyl} the magnetic field parameter is chosen to be $\beta=0.2\lambda$, while $\ell$ is varied. We see that increasing $\ell$ increases the height of the embedding geometry. In Fig.~\ref{fig_embedding_Varybeta} the cosmological constant parameter is $\ell=100\lambda$ while $\beta$ is varied. In this case, increasing $\beta$ has the effect of reducing the width of the embedding geometry.}
 \label{fig_embedding}
\end{figure}

Another way to visually depict the geometry is by an embedding diagram. In particular, we embed the $(r,\psi)$-section of the metric, $\frac{\dif r^2}{f}+f\dif\psi^2$, into a spatial Euclidean 3-space $\mathbb{R}^3$. In cylindrical coordinates, its metric is $\dif s^2_{\mathbb{R}^3}=\dif Z^2+\dif R^2+R^2\dif\varphi^2$. Writing $Z=Z(r)$ and $R=R(r)$, the Euclidean metric restricted to the embedding is
\begin{align}
 \dif s^2_2=\brac{Z'^2+R'^2}\dif r^2+R^2\dif\psi^2/C^2,
\end{align}
where $\varphi=\psi/C$, $R=\sqrt{f}$ and $Z'=\sqrt{\frac{1-f'^2/4}{f}}$. Because of the square root in the expression for $Z'$, the solution is only embeddable if $f'^2<4$. The constant $C$ is taken to be equal to $2\sbrac{(d-3)\beta^2/\lambda^{2d-5} - (d-1)\lambda/\ell}^{-1}$ so that the periodicity of $\varphi$ can be taken to be $2\pi$ and the embedding surface in $\mathbb{R}^3$ is smooth at the point $r=\lambda$, reflecting the absence of conical singularity there. We find that only Case A parameters satisfy this condition. Solving this differential equation for $Z$ gives a curve $(R(r),Z(r),\varphi=0)$ in $\mathbb{R}^3$. The $(r,\psi)$-section as an embedded surface is obtained as a surface of revolution of the curve about the $Z$-axis. In Fig.~\ref{fig_embedding}, we show these curves for Case A solutions with $d=4$, $\beta=0.1\lambda$, and various $\ell$. In $\mathbb{R}^3$ as depicted here, the geometry starts at $Z=0$, corresponding to the $r=\lambda$ tip, where it appears smooth as we have removed the conical singularity there. The upper tip corresponds to the $r=r_0$, and carries the aforementioned conical deficit.

It is interesting to note that the values of $\ell$ and $\beta$ affects the heights and widths of the embeddings, respectively. In Fig.~\ref{fig_embedding_Varyl}, we show the embeddings for fixed $\beta$ while varying $\ell$. We see that as $\ell$ is increased, the shape becomes more elongated, and the sharp upper tip reflects the conical singularity at $r=\lambda$, which is more severe as $\ell$ increases. In the limit $\ell\rightarrow\infty$, the upper tip is sent to infinity and we obtain the `tall-necked vase' described in \cite{Thorne:1965}. On the other hand we fix $\ell$ and vary $\beta$ in Fig.~\ref{fig_embedding_Varybeta}. Here, the heights remain roughly the same while the widths decreases with increasing $\beta$.

\subsection{Absence of horizons and de Sitter limit}

 {We have now seen that the domain for $r$ is finite and does not include zero if the cosmological constant is positive. A particular consequence of this is that the norm-squared of the time-like Killing vector $\xi^\mu=\partial_t^\mu$ is proportional to $ -r^2/\lambda^2$, hence is nowhere vanishing, and there are no horizons in this spacetime. Furthermore, this spacetime does not contain de Sitter space as any limit nor asymptote to it. With these considerations, it is perhaps inappropriate to call this solution `\emph{Melvin--de Sitter}'. Instead, in this paper we simply refer to it as a `\emph{Melvin spacetime with positive cosmological constant}'.}

 {In the zero and negative cosmological constant cases, switching off the magnetic field reduces the spacetime to Minkowski or Anti-de Sitter, respectively. Contrarily, the positive cosmological constant case does not reduce the spacetime to de Sitter. In Sec.~\ref{subsec_ParamSpace}, we have seen that setting $\beta=0$ leaves behind a naked curvature singularity. We can try to see what happens if $\beta$ is taken to zero as a limit by checking the behaviour of the function $f$, such as in Fig.~\ref{fig_froots_fixedl_lambdaleft}. Loosely speaking, the value of $f$ at a particlular $r$ indicates the size of their respective constant-$r$ circles. For $\beta\neq0$, $f$ always has a maximum between two roots. Decreasing $\beta$ brings the smaller root closer to zero and the maximum of $f$ larger. In the limit $\beta\rightarrow\infty$, $f$ diverges at zero, along with the Kretschmann scalar. So switching off the magnetic field results in decompactification of the spacetime and leaving behind a curvature singularity. This singular limit is clearly not de Sitter and still horizonless, though the norm of the time-like Killing vector vanishes at the $r=0$ singularity. Unlike the $\Lambda=0$ and $\Lambda<0$ counterparts, turning off the field does not reduce to their respective `pure' vacuum solutions.}

\section{Flux} \label{sec_flux}

 {It is clear that the magnetic field plays an important role in shaping the geometry of the Melvin spacetime, seeing that many of the geometrical properties in the previous section depend on $\beta$. Conversely, the geometry of the spacetime may in turn affect the configuration of the field. Therefore in this section we turn to study the magnetic field itself. The gauge potential $A$ is described by \Eqref{GaugePotential}, from which we calculate the field components by $F=\dif A$. This, however, is coordinate-dependent and it should be more useful to consider the flux which is an invariant quantity. Below we will compute the flux and see what it tells us about the distribution of magnetic fields in the Melvin geometry. From this we may also see how the repulsive nature of a positive $\Lambda$ which provides the tendency to disperse the fields.}

Let $C$ be a curve of constant-$r$ and $x^a$ centred at $r=\lambda$. Then the flux through the region bounded by $C$ is
\begin{align}
 \Phi(r)=\int_{\mathrm{int}(C)} F=\int_CA=\Delta\psi\sqrt{\frac{d-2}{2(d-3)}}\beta\left|\frac{1}{\lambda^{d-3}}-\frac{1}{r^{d-3}} \right|,
\end{align}
Let us first suppose that $\lambda<r_0$. That is, $\beta\ell>\eta$. This case contains the limit $\ell\rightarrow0$ to $\Lambda=0$ for fixed $\beta$. For small circles, we write $r=\lambda+\epsilon$, and the flux is approximately given by
\begin{align}
 \Phi\simeq\Delta\psi\sqrt{\frac{d-2}{2(d-3)}}\;\beta\frac{d-3}{\lambda^{d-2}}\epsilon+\mathcal{O}(\epsilon^2).
\end{align}
We can compare the results from previous literature by using Eq.~\Eqref{AstorinoCoord} to see that $r-\lambda=\epsilon=\frac{1}{4}B^2\rho^2$. Furthermore using Eq.~\Eqref{Btransform} to express $\beta$ in terms of $B$, we find that
\begin{align}
 \Phi\simeq\sqrt{\half(d-2)(d-3)}\frac{\pi\rho^2B}{1-\frac{(d-1)(d-3)}{\ell^2B}}+\mathcal{O}(\rho^4),
\end{align}
so if $\ell B$ is large and $d=4$, we recover
\begin{align}
 \Phi\simeq \pi\rho^2B, \label{non_GR_flux}
\end{align}
in agreement with the $\Lambda<0$ and $\Lambda=0$ cases \cite{Kastor:2020wsm}. This result simply reflects the fact that close to the $r=\lambda$ pole, the magnetic field is approximately uniform on an approximately flat background. Hence the flux contained is  the same as a non-gravitational uniform magnetic field of strength $B$ contained in the area $\pi\rho^2$ of the circle of radius $\rho$. This approximation required the condition $\frac{(d-1)(d-3)}{B^2\ell^2}\ll 1$, which tells us that $\ell$ must be large enough such that the spacetime curvature due to the cosmological constant is not noticeable in that region.

\begin{figure}
 \centering
 \begin{subfigure}[b]{0.49\textwidth}
    \centering
    \includegraphics[width=\textwidth]{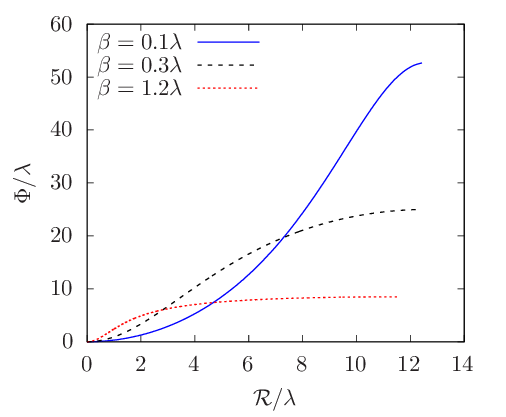}
    \caption{$\ell=10\lambda$.}
    \label{fig_flux_FluxR_l10}
  \end{subfigure}
  \begin{subfigure}[b]{0.49\textwidth}
    \centering
    \includegraphics[width=\textwidth]{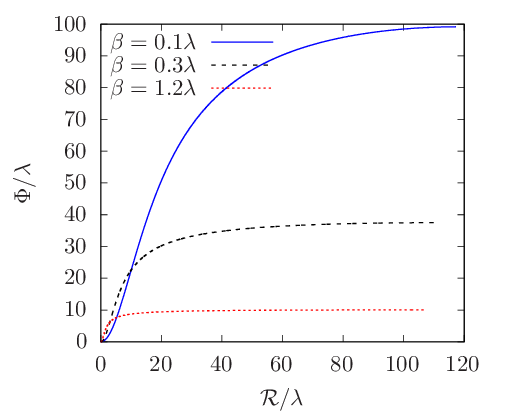}
    \caption{$\ell=100\lambda$.}
    \label{fig_flux_FluxR_l100}
  \end{subfigure}
  \caption{Plots of flux $\Phi$ contained in circles of proper radius $\mathcal{R}$ centred at $r=\lambda$.}
  \label{fig_flux_FluxR}
\end{figure}

For circles further away from $\lambda$, instead of the coordinate $r$, it might be more meaningful to consider the proper radius of the circle $\mathcal{R}$ (defined in Eq.~\Eqref{define_Rcal}).
In Fig.~\ref{fig_flux_FluxR} we plot $\Phi(r)$ against $\mathcal{R}(r)$ for $d=4$. For small $\mathcal{R}\sim 0$ (equivalently $r\sim\lambda$,) the flux was shown to behave like Eq.~\Eqref{non_GR_flux}. So the solutions with larger $\beta$ (equivalently, larger $B$) grows faster. However those with larger $\beta$ also has their the growth plateau more quickly as the circle radii increases.  {Because of the compactness of the $r$ direction, the notion of a `fluxtube radius' used in \cite{Kastor:2020wsm} may not be well-defined for the present solution. Nevertheless, we might interpret the approximate radii where the plateau starts as representing the fluxtube radius, as `most' of the flux is contained within this radius. (Provided that we measure it from the $r=\lambda$ pole.) This radius gets smaller with large and increasing $\beta$. In other words, this is a similar behaviour as the zero and negative cosmological constant cases \cite{Kastor:2020wsm}, where fluxtubes with strong magnetic fields are narrow.}

\begin{figure}
 \centering
 \includegraphics{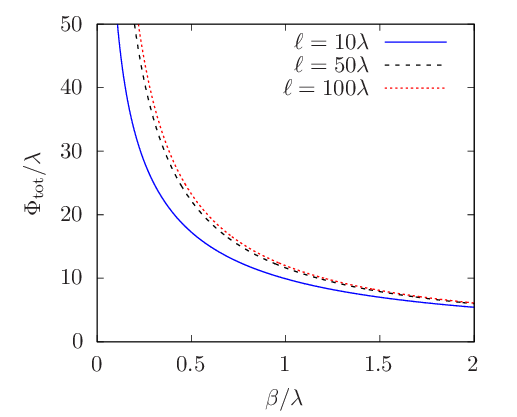}
 \caption{Plots of total flux $\Phi_{\mathrm{tot}}$ against $\beta$.}
 \label{fig_flux_TotalFlux}
\end{figure}

The total flux contained in the compact domain $\lambda\leq r\leq r_0$ or $r_0\leq r\leq \lambda$ is given by $\Phi_{\mathrm{tot}}=\Phi(r_0)$. In Fig.~\ref{fig_flux_TotalFlux} we plotted the total flux against the field strength parameter $\beta$, for various $\ell$. We find the behaviour similar to the $\Lambda<0$ case where the dependence on $\Phi_{\mathrm{tot}}$ on $\beta$ is roughly inversely proportional \cite{Kastor:2020wsm}. In fact, from the expression
\begin{align}
 \Phi_{\mathrm{tot}}=\Delta\psi\sqrt{\frac{d-2}{2(d-3)}}\beta\left|\frac{1}{\lambda^{d-3}}-\frac{1}{r_0^{d-3}} \right|,
\end{align}
we notice that from Eq.~\Eqref{set_delta_psi}, $\Delta\psi\sim 1/\beta^2$. As $\beta$ becomes large, $r_0$ increases. (This can be seen, e.g., from the numerical example in Fig.~\ref{fig_froots_fixedl_lambdaright}.) Therefore the contribution of the term $-1/r_0^{d-3}$ becomes negligible and $\Phi_{\mathrm{tot}}\sim1/\beta$ for large $\beta$. For $\beta\rightarrow0$, the total flux diverges.  {One can then compare this with the $\Lambda=0$ and $\Lambda<0$ cases studied previously. In particular, it was argued in \cite{Kastor:2020wsm} that a negative cosmological constant serves as a confining box, containing the flux within the order of the AdS radius, and no such confinement occurs for the $\Lambda=0$ case, where weak magnetic fields are being spread out over a large region.

It is then natural to expect that the $\Lambda>0$ case is the opposite to $\Lambda<0$. That is, instead of a confining box, we have a repulsive `\emph{anti-box}'\footnote{We gratefully acknowledge the anonymous referee for suggesting this point and introducing this terminology.} which spreads out the flux even more than the $\Lambda=0$ case. We can see this, for instance, in Fig.~\ref{fig_flux_FluxR_l10}, where the solution with the weaker field are being spread out over a larger radius, as the flux pleateaus much later for smaller $\beta$. In Fig.~\ref{fig_flux_TotalFlux}, we plot the total flux $\Phi_{\mathrm{tot}}$ against the field strength $\beta$. We see that the total flux diverges in the limit $\beta\rightarrow0$, similar to the $\Lambda=0$ case. The flux diverges more strongly for larger positive $\Lambda$ (i.e., smaller $\ell$), exibiting the `anti-box' behaviour of dispersing the field more strongly than the $\Lambda=0$ case.}

\section{Geodesics} \label{sec_geodesics}

 {In this section we investigate the dynamics of a time-like or null particle moving in the Melvin spacetime with positive cosmological constant. Such an analysis will be useful since the intuitive effects of gravity in a given spacetime can be revealed by looking at the motion of particles within it.}

The trajectories of such particles follow geodesics, which we describe here by curves $x^\mu(\tau)$, where $\tau$ is an appropriate affine parametrisation. The Lagrangian for a test particle undergoing geodesic motion is
\begin{align}
 \mathcal{L}=\half\sbrac{\frac{r^2}{\lambda^2}\brac{-\dot{t}^2+\dot{x}_1^2+\ldots+\dot{x}_{d-3}^2}+\frac{\dot{r}^2}{f}+f\dot{\psi}^2},
\end{align}
where over-dots denote derivatives with respect to $\tau$. The coordinate $r$ is the only one explicitly appearing in the Lagrangian. Hence we have conserved momenta associated to each of the other directions namely,
\begin{align}
 E=\frac{r^2}{\lambda^2}\dot{t},\quad p_j=\frac{r^2}{\lambda^2}\dot{x}_j,\quad L=f\dot{\psi},\label{FirstIntegrals}
\end{align}
where $E$, $L$, and $p_j$, $j=1,\ldots,d-3$ are regarded as the particle's energy, angular momentum, and linear momenta, respectively. To account for both time-like and null geodesics, the parametrisation of $\tau$ is fixed such that $g_{\mu\nu}\dot{x}^\mu\dot{x}^\nu=\epsilon$, where explicitly it is written as
\begin{align*}
 \frac{r^2}{\lambda^2}\brac{-\dot{t}^2+\dot{x}_1^2+\ldots+\dot{x}^2_{d-3}}+\frac{\dot{r}^2}{f}+f\dot{\psi}^2=\epsilon=\left\{\begin{array}{cc}
  -1, & \mbox{time-like},\\
   0, & \mbox{null}.
  \end{array}\right.
\end{align*}
Using Eq.~\Eqref{FirstIntegrals}, we get an equation of motion for $r$,
\begin{align}
 \dot{r}^2&=\frac{\lambda^2f}{r^2}\brac{E^2-P^2}-L^2+\epsilon f,\label{Ueff}
\end{align}
where $P^2=p_1^2+\ldots+p_{d-3}^2$. However, we can always transform to a frame where the $\dot{x}_j$, and hence $p_j$ is zero. Therefore in the following we will set $P=0$ without loss of generality.

\textbf{Time-like geodesics.} We first consider time-like geodesics, where $\epsilon=-1$. In this case Eq.~\Eqref{Ueff} is $\dot{r}^2+U_{\mathrm{eff}}=0$, where
\begin{align}
 U_{\mathrm{eff}}=L^2-\sbrac{\frac{\lambda^2}{r^2}\brac{E^2-P^2}-1}f.
\end{align}
Clearly, the allowed domain of particle motion is where $U_{\mathrm{eff}}\leq0$. If $L=0$, the domain $U_{\mathrm{eff}}\geq0$ coincides with $f\leq0$, which is between the roots $\lambda$ and $r_0$. By numerical exploration, we find that, for fixed $E$, the domain of allowed $r$ becomes narrower as $L$ is increased. The domain also becomes narrower by decreasing $E$ with fixed $L$. There is a critical point for which the $r$-domain shrinks to a point, in which the geodesic is a constant-$r$ circular orbit.

A circular orbit of radius $r=r_c$ can be obtained by solving $U_{\mathrm{eff}}=U'_{\mathrm{eff}}=0$ to obtain the required energy $E_c$ and and angular momentum $L_c$ as expressions parametrised in terms of $r=r_c$. Explicitly, they are
\begin{align}
 L_c^2=\frac{2f(r_c)^2}{r_c\brac{f'(r_c)-\frac{2}{r_c}f(r_c)}},\quad E_c^2=\frac{r_cf'(r_c)}{\lambda^2\brac{f'(r_c)-\frac{2}{r_c}f(r_c)}}, \label{CircularTimelikeEL}
\end{align}
and we see that $U_{\mathrm{eff}}''(r_c)>0$, so all circular orbits are stable. Note that the squared quantity $L_c^2$ is assured to be positive if $f'(r_c)>\frac{2}{r_c}f(r_c)$, or
\begin{align}
 r_c<\lambda\brac{\frac{2(d-2)}{d-1}\frac{\beta^2}{\beta^2+\frac{\lambda^{2(d-2)}}{\ell^2}}}^{\frac{1}{d-3}}, \label{CircularTimelikeBound}
\end{align}
giving the upper bound of the radii of circular orbits. In particular, for $d=4$ and $\ell\rightarrow\infty$, this bound is $r_c<\frac{4}{3}\lambda$, recovering Melvin and Wallingford's result \cite{Melvin:1965zz}.\footnote{Melvin and Wallingford's cylindrical coordinates is related to our present $r$ by $\frac{r}{\lambda}=1+\rho^2$. Therefore $\frac{r}{\lambda}<\frac{4}{3}$ is equivalent to $\rho<\frac{1}{\sqrt{3}}$.} (See also \cite{Bini:2022xzk}.)

\begin{figure}
  \begin{subfigure}[b]{0.49\textwidth}
    \centering
    \includegraphics[width=\textwidth]{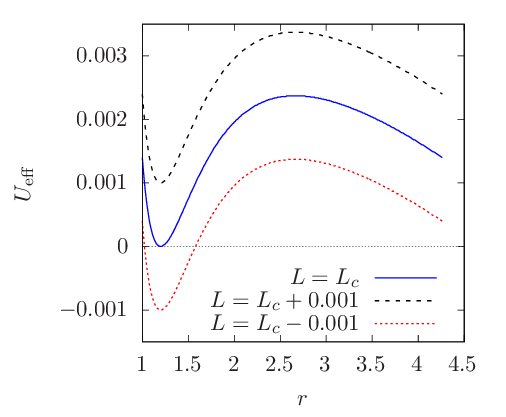}
    \caption{$E=E_c$.}
    \label{fig_UeffTimelikeYTVaryL}
  \end{subfigure}
  \begin{subfigure}[b]{0.49\textwidth}
    \centering
    \includegraphics[width=\textwidth]{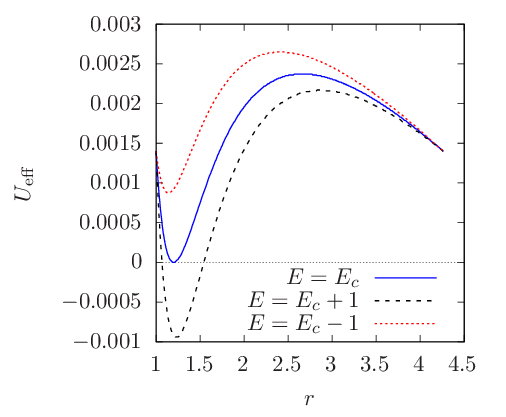}
    \caption{$L=L_c$.}
    \label{fig_UeffTimelikeYTVaryE}
  \end{subfigure}
  \caption{Graphs of effective potential $U_{\mathrm{eff}}$ for $d=4$, $\ell=100\lambda$, and $\beta=0.1\lambda$. Here $L_c=0.03736$ and $E_c=1.7185$ are the energy and angular momentum for a circular orbit of radius $r_c=1.2\lambda$.}
  \label{fig_UeffTimelikeYT}
\end{figure}

Lowering the angular momentum below $L_c$ or raising the energy above $E_c$ will result in a $U_{\mathrm{eff}}$ that is negative in some finite domain $r_-<r<r_+$, and the geodesic oscillates in a bound orbit between $r_-$ and $r_+$. As an example, in Fig.~\ref{fig_UeffTimelikeYT}, we plot $U_{\mathrm{eff}}$ for $d=4$, $\ell=100\lambda$, and $\beta=0.1\lambda$, with a circular orbit chosen to be $r=1.2\lambda(<\frac{4}{3}\lambda)$. In this case the angular momentum and energy for the circular orbit are, up to five significant figures, $L_c=0.037361$ and $E_c=1.7185$.

\textbf{Null geodesics.} In the case of null geodesics which describes the trajectories of photons, we have $\epsilon=0$, and Eq.~\Eqref{Ueff} becomes $\dot{r}^2+V_{\mathrm{eff}}=0$, where
\begin{align}
 V_{\mathrm{eff}}&=L^2-\frac{\lambda^2}{r^2}\brac{E^2-P^2}f.
\end{align}
As before, orbits exists in the domain for which $V_{\mathrm{eff}}\leq0$. For $L=0$, this domain coincides with the domain for $f\geq0$, meaning particles of zero angular momentum can reach the poles $\lambda$ and $r_0$. For non-zero $L$, this domain is smaller than the domain for $f\geq0$, hence photons of non-zero angular momentum cannot reach the poles.

For circular null geodesics, we solve $V_{\mathrm{eff}}=V_{\mathrm{eff}}'=0$, which lead to
\begin{align}
 f'(r_c)=\frac{2}{r_c}f(r_c)\quad\leftrightarrow\quad r_c=\lambda\brac{\frac{2(d-2)}{d-1}\frac{\beta^2}{\beta^2+\frac{\lambda^{2(d-2)}}{\ell^2}}}^{\frac{1}{d-3}}. \label{CircularNull_rc}
\end{align}
Hence circular photon orbits only occur at exactly this radius. Note that this is the limit where the inequality \Eqref{CircularTimelikeBound} of time-like circular orbits is saturated, or equivalently, this is the limit $L_c^2,E_c^2\rightarrow\infty$ in Eq.~\Eqref{CircularTimelikeEL}. This reflects the fact that the circular photon orbits is the limit of circular time-like orbits with infinite energy and angular momentum. In the zero cosmological constant case, this radius is independent of the magnetic field strength, as can be seen by taking $d=4$ and $\ell\rightarrow\infty$ in Eq.~\Eqref{CircularNull_rc} leading to $r_c=\frac{4}{3}\lambda$, recovering the result of Melvin and Wallingford which was $\rho=\frac{1}{\sqrt{3}}$ \cite{Melvin:1965zz}.

\section{Conclusion} \label{sec_conclusion}

In this paper we have studied the geometry of the Melvin universe with positive cosmological constant. The spacetime has the topology of the form $\mathbb{R}^{1,d-3}\times S^2$. In particular, the `radial' coordinate $r$ is compact, unlike its $\Lambda\leq0$ counterparts. The solution carries a conical singularity, specifically a conical deficit if $\beta\ell>\sqrt{\frac{d-1}{d-3}}\lambda^{d-2}$ and a conical excess if $\beta\ell<\sqrt{\frac{d-1}{d-3}}\lambda^{d-2}$. The limit $\beta\ell\rightarrow\sqrt{\frac{d-1}{d-3}}\lambda^{d-2}$, along with an appropriate scaling of the $r$-coordinate, is shown to be a Freund--Rubin-type flux compactification model.

Spacetime geodesics representing the motion of time-like and null particles was also studied. We have seen that all orbits are bounded, in similarity to the $\Lambda=0$ spacetime. But the present case is perhaps even less surprising due to the compact $S^2$ topology. The motion here is akin to motion confined on a surface of a sphere, and particles with non-zero angular momentum (defined with respect to an axis at the poles) are unable to reach the poles.

Most of the analysis was done in a radial coordinate $r$ that is related to the usual Melvin `cylindrical' coordinate $\rho$ by $\frac{r}{\lambda}=\brac{1+\frac{1}{4}B^2\rho^2}^{1/(d-3)}$ where $\lambda$ is an arbitrary length scale. In the $r$-coordinates, the Melvin spacetime is a double-Wick-rotated version of the charged black hole with planar horizons. This viewpoint may suggest that a natural extension to the present set of Melvin solutions is to consider double-Wick rotations of charged black holes with \emph{spherical} or \emph{hyperbolic} horizons. Indeed, the former case would yield solutions of the form $S^2\times\mathrm{dS}_{d-2}$ upon the Wick rotations. While this solution may not be of `Melvin-type', they have been viewed as models of $d=6$-dimensional braneworld gravity \cite{Mukohyama:2005yw,Kinoshita:2006eld,Kinoshita:2007ci}. 

\section*{Acknowledgments}
 
Y.-K.~L is supported by Xiamen University Malaysia Research Fund (Grant no. XMUMRF/ 2021-C8/IPHY/0001).

\bibliographystyle{dS-Melvin}

\bibliography{dS-Melvin}

\end{document}